\documentclass[a4paper, 10pt, unnumberedsections, twoside]{LTJournalArticle}

\usepackage[acronym]{glossaries}
\usepackage[detect-all=true,per-mode=symbol,per-symbol =/]{siunitx}
\usepackage[capitalise]{cleveref}
\usepackage{tablefootnote}
\usepackage{threeparttable}
\usepackage{booktabs}
\usepackage{xcolor}
\usepackage{subcaption}
\usepackage{makecell}
\usepackage{tikz}
\usepackage{amssymb}

\DeclareSIUnit{\x}{\!\ensuremath{\times}}
\DeclareSIUnit\bit{b}
\DeclareSIUnit\gateeq{GE}
\DeclareSIUnit\pixel{px}
\DeclareSIUnit\dpi{dpi}
\DeclareSIUnit\bit{bit}
\newacronym{api}{API}{application programming interface}
\newacronym{asic}{ASIC}{application-specific integrated circuit}
\newacronym{drc}{DRC}{design rule check}
\newacronym{gdsii}{GDSII}{Graphic Data System II}
\newacronym{ic}{IC}{integrated circuit}
\newacronym{oseda}{OSEDA}{open-source electronic design automation}

\newcommand{\riscv}{\mbox{RISC-V}}

\newcommand{\artistic}{ArtistIC}
\newcommand{\klayout}{KLayout}
\newcommand{\imagemagick}{ImageMagick}

\newcommand*\circnum[1]{\tikz[baseline=(char.base)]{%
            \node[white,shape=circle,fill=black,draw,inner sep=1pt] (char) {\color{white}\sffamily #1};}}

\runninghead{}
\footertext{}
\title{{\artistic}: An Open-Source Toolchain for Top-Metal IC Art and Ultra-High-Fidelity GDSII Renders}

\author{%
    Thomas Benz\textsuperscript{1}\thanks{~Corresponding author: \href{mailto:tbenz@iis.ee.ethz.ch}{\tt tbenz@iis.ee.ethz.ch}}, \
    Paul Scheffler\textsuperscript{1}, \
    Nils Wistoff\textsuperscript{1},\\ \
    Philippe Sauter\textsuperscript{1}, \
    Beat Muheim\textsuperscript{1}, \
    Luca Benini\textsuperscript{1,2}
}

\newcommand{\x}{$\times$}

\date{
    \vspace{-0.32em}
    \footnotesize\textsuperscript{\textbf{1}}Integrated Systems Laboratory, ETH Zurich \\
    \footnotesize\textsuperscript{\textbf{2}}Department of Electrical, Electronic, and Information Engineering, University of Bologna
    \vspace{-0.32em}
}

\renewcommand{\maketitlehookd}{%
	\begin{abstract}
		\noindent Open-source projects require outreach material to grow their community, secure funds, and strengthen their influence.
        Numbers, specifications, and facts alone are intangible to uninvolved people; using a clear brand and appealing visual material is thus ample to reach a broad audience.
        This is especially true for \glspl{asic} during the early stages of the development cycle without running prototype systems.
        This work presents {\artistic}, an open-source framework to brand \glspl{asic} with top-metal art and to render \glsunset{gdsii}\gls{gdsii} layouts with ultra-high fidelity reaching render densities below \SI{25}{\nano\metre\per\pixel} and gigapixels-scale resolutions.

	\end{abstract}
}

\begin{document}

\maketitle
\glsresetall

\section{Introduction}

Recent advances in \gls{oseda} spawned countless \gls{asic} projects developed by industry, research, and hobbyists.
Even though the content and organization of these projects could not be more diverse, they share a common requirement to present their project using outreach material to gain attention, share resources and results, and secure funding.
Especially in the early design phases of projects, sharing the \gls{asic} under design can prove very difficult as no running prototype can be shown. 

Methodologies have been established to render and visualize layout (\glsunset{gdsii}\gls{gdsii}) files, but they are limited either in scope or fidelity.
The \emph{TinyTapeout} project~\cite{venn2024tiny} uses a {3D}-\gls{gdsii} viewer to visualize their layouts directly within a web browser~\cite{mbalestrini2022ttgdsviewer}.
Proving an excellent tool to visualize tiny designs on the standard-cell level in {3D}, it is not suited to render research or industry-grade chips on a poster scale.
{\klayout}~\cite{kofferlein2020klayout} has established itself as the default open-source \gls{gdsii} viewer featuring a powerful scripting \gls{api}, which can be used to export high-resolution ($\lesssim$ \SI{250}{\mega\pixel}) renders of the current view~\cite{kofferlein2019screenshot}.
Our experiments show that a much higher resolution is required to capture the intricate details of the lower-level metalization layers of research-scale~\cite{sauter2024insights, scheffler2025occamy} chips.
Furthermore, {\klayout} does not provide a transparency option for individual layers, leading to upper metal levels covering up lower-level details.

Adding top-metal artwork is a well-established way of branding the design files and the fabricated silicon~\cite{goldstein2002secret}.
To our knowledge, no open-source tool exists to translate and embed artwork into \gls{gdsii} layouts.

This work proposes {\artistic}, an open-source framework\footnote{\url{github.com/pulp-platform/artistic}} to translate and insert top-metal \gls{asic} art into \gls{gdsii} layout files and create artistic ultra-high-fidelity renders thereof.
In particular, we present the following contributions:
\begin{itemize}
    \item A \emph{Gdspy}-based~\cite{2024gdspy} script translating and inserting \glsunset{drc}\gls{drc}-clean top-metal artworks into \gls{gdsii} files.
    \item A tile-based image rendering methodology supporting arbitrary high resolutions through tiling and capable of individual layer transparencies.
    \item A case study presenting insights from analyzing the layout renders of two {\riscv} \glspl{asic}~\cite{sauter2024insights, scheffler2025occamy}.
\end{itemize}

\section{Toolflow}

\begin{figure}
\centering%
    \includegraphics[width=1\linewidth]{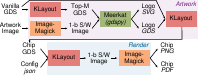}%
    \caption{The {\artistic} toolflow.}
    \label{fig:toolflow}
\end{figure}

{\artistic}'s toolflow, presented in \Cref{fig:toolflow}, accepts a \gls{drc}-clean \emph{vanilla} \gls{gdsii} file and a lossless artwork image as primary inputs.
In the first step, {\klayout} is used to export the top-metal layer from the \gls{gdsii} file and the logo is translated into a 1-\si{\bit} \emph{b/w} image.
\emph{Meerkat}, a \emph{Gdspy}-based script, converts the \emph{b/w} image to \emph{tetromino}-like shapes following \gls{drc} and density rules as well as flowing them around existing top-metal structures.
\emph{Meerkat} exports a vector graphic image and a \gls{gdsii} file containing the logo only; {\klayout} is then used to merge the vanilla and logo \gls{gdsii} files.

A \emph{json} configuration file specifies the resulting image size, render resolution, layer colors and transparencies, and the used layer stack. 
Using a maximum individual tile size of around \SI{250}{\mega\pixel}, we use {\klayout} to export each layer of the \gls{asic} as \emph{b/w} image tiles.
The tiles can be rendered at a higher resolution than the final image size to keep fine layout details while ensuring a manageable image file size. 
In the last step, \emph{\imagemagick} is used to color the layer tiles, merge the colored layers using the specified transparencies, and resize and merge the tiles to a complete image, which is then embedded into a \emph{PDF} container for printing.

\section{Results}

\begin{figure}
    \begin{subcaptionblock}{0.47\linewidth}%
        \centering%
        \includegraphics[width=\linewidth]{fig-02.png}%
        \vspace{-0.4em}%
        \caption{Klayout render}%
        \vspace{0.25em}%
        \label{fig:klayout_render}%
        \vfill%
        \centering%
        \includegraphics[width=\linewidth]{fig-03.png}%
        \vspace{-0.3em}%
        \caption{{\artistic} render}%
        \label{fig:artistic_render}%
    \end{subcaptionblock}\hfill
    \begin{subcaptionblock}{0.52\linewidth}
        \centering%
        \includegraphics[width=\linewidth]{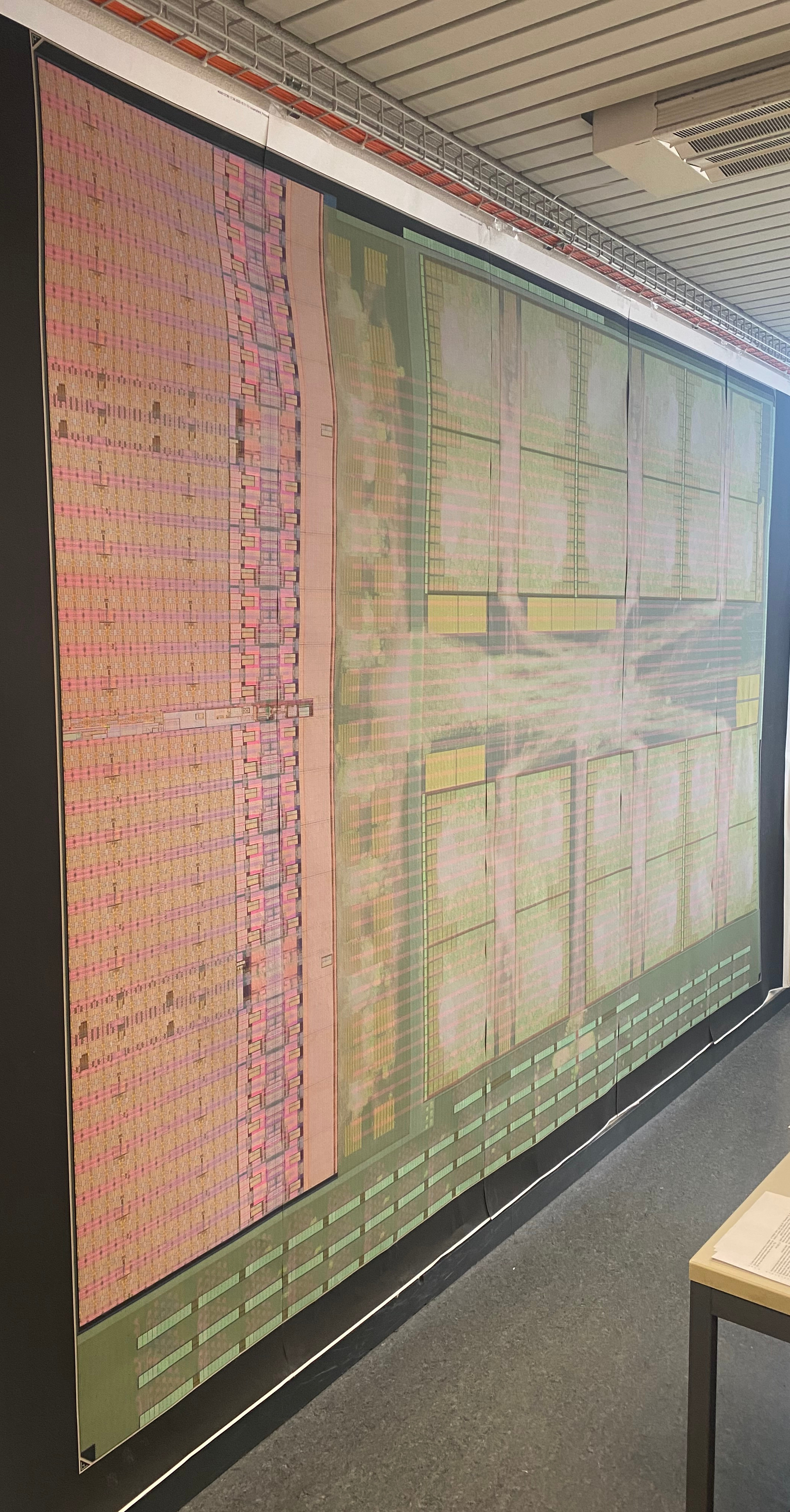}%
        \vspace{-0.3em}%
        \caption{Wall-spanning poster}%
        \label{fig:artistic_poster}%
    \end{subcaptionblock}\hfill

    \caption{\gls{gdsii} renders}
    \label{fig:artistic}
\end{figure}

\Cref{fig:klayout_render} shows the \emph{vanilla} \gls{gdsii} file~\cite{pulpplatform2024mlem} rendered with {\klayout}~\cite{kofferlein2020klayout, kofferlein2019screenshot} and \Cref{fig:artistic_render} with {\artistic} including a generated top-level metal artwork.
\Cref{tab:runtime} presents {\artistic}'s runtimes for two open-source \glspl{asic}.
\Cref{fig:artistic_poster} displays a 2.3 by \SI{4.2}{\metre} poster of one Occamy die~\cite{scheffler2025occamy} rendered at \SI{57}{\giga\pixel} (\SI{44}{\nano\metre\per\pixel}) and printed at \SI{1600}{\dpi} demonstrating the scalability of {\artistic}.

\newcommand{\dl}[2]{\makecell[cc]{#1 \\ #2}}

\begin{table}
    \setlength{\tabcolsep}{4pt}
    \centering
    \scriptsize{%
        \centering
        \caption{%
            {\artistic} runtime of open-source {\riscv} \glspl{asic}.%
        }%
        \label{tab:runtime}
        \renewcommand*{\arraystretch}{0.95}
        \begin{threeparttable}
            \begin{tabular}{cccccc} \toprule
                \textbf{Chip} &
                \dl{\textbf{Chip}}{\textbf{Size}} &
                \dl{\textbf{Render}}{\textbf{Res.}} &
                \dl{\textbf{Print}}{\textbf{Res.}} &
                \dl{\textbf{Print}}{\textbf{Size}} &
                \dl{\textbf{Run-}}{\textbf{time}} \\

                \midrule

                \textit{Mlem/Croc~\cite{pulpplatform2024mlem}} &
                \SI{5}{\milli\metre\squared} &
                \SI{25}{\nano\metre\per\pixel} &
                \SI{2}{\giga\pixel} &
                A1~\tnote{b} &
                \SI{0.9}{\hour}~\tnote{a} \\

                \textit{Basilisk~\cite{sauter2024insights}} &
                \SI{35}{\milli\metre\squared} &
                \SI{25}{\nano\metre\per\pixel} &
                \SI{55}{\giga\pixel} &
                2x\SI{2}{m}~\tnote{b} &
                \SI{6.1}{\hour}~\tnote{a} \\

                \bottomrule
                
            \end{tabular}

            \begin{tablenotes}[para, flushleft]
                \item[a] \SI{2.5}{\GHz} Xeon E5-2670
                \item[b] Target: \SI{2400}{\dpi}
            \end{tablenotes}
        \end{threeparttable}
    }
\end{table}

\section{Case Study: RISC-V SoCs}

\begin{figure}
    \begin{subcaptionblock}{0.45\linewidth}
        \centering%
        \includegraphics[width=\linewidth]{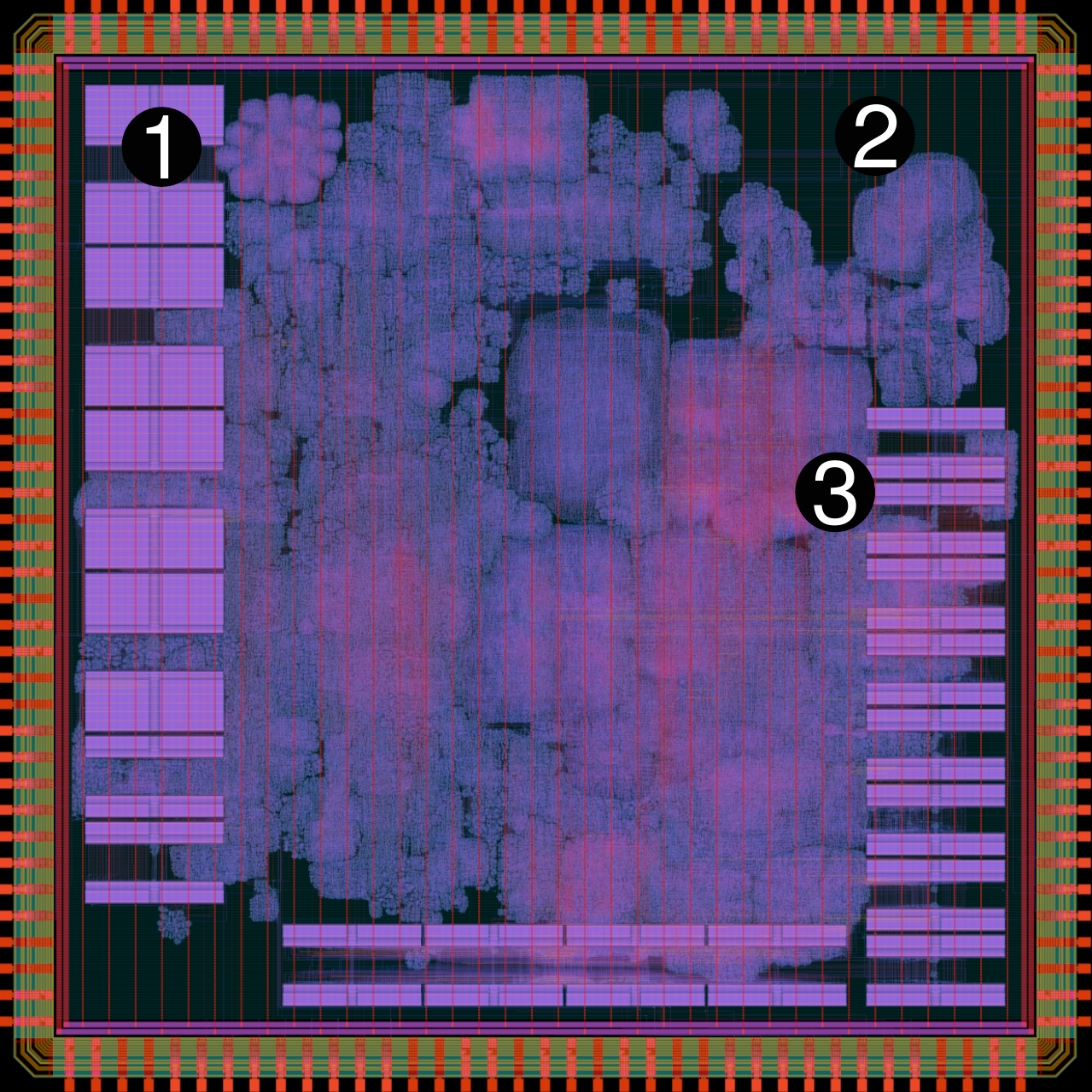}%
        \vspace{-0.3em}%
        \caption{Linux-capable SoC~\cite{sauter2024insights}}%
        \label{fig:artistic_iguana}%
    \end{subcaptionblock}\hfill
    \begin{subcaptionblock}{0.508\linewidth}%
        \centering%
        \includegraphics[width=\linewidth]{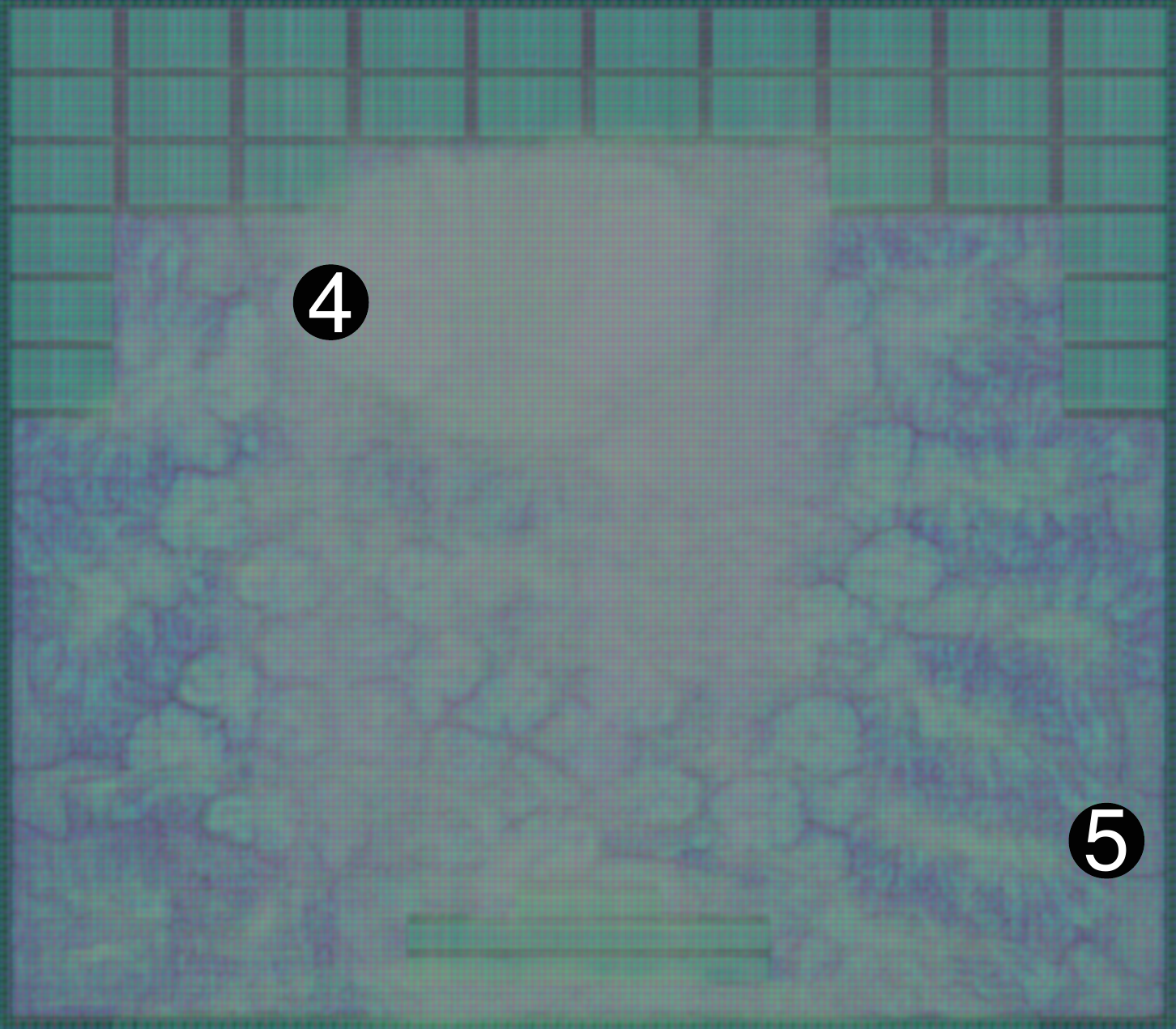}%
        \vspace{-0.3em}%
        \caption{Occamy cluster~\cite{scheffler2025occamy}}%
        \label{fig:artistic_cluster}%
    \end{subcaptionblock}\hfill
    \begin{subcaptionblock}{\linewidth}
        \centering%
        \includegraphics[width=\linewidth]{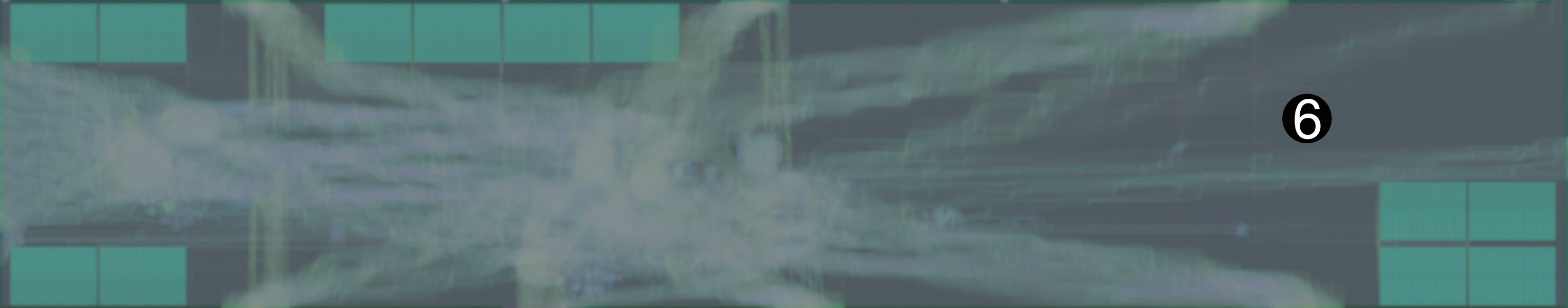}%
        \vspace{-0.3em}%
        \caption{Occamy main interconnect~\cite{scheffler2025occamy}}%
        \label{fig:artistic_ico}%
    \end{subcaptionblock}\hfill

    \caption{Insights gained through {\artistic}.}
    \label{fig:case}
\end{figure}

{\artistic} can directly be used in scientific publications to analyze, evaluate, and compare \gls{asic} layouts.
\Cref{fig:case} presents three high-fidelity layout renders with annotations of exemplary insights gained; for a more detailed analysis, we refer to our peer-reviewed works on these systems~\cite{sauter2024insights, scheffler2025occamy}.
In \Cref{fig:artistic_iguana}, we see \circnum{1} the bootrom, which is closely interconnected and thus densely packed, \circnum{2} empty space, and \circnum{3} an area of high connectivity; the routing uses higher top metals (rendered in red).
In \Cref{fig:artistic_cluster}, \circnum{4} highlights the high routing effort of the \emph{logarithmic interconnect} and \circnum{5} the \emph{pipeline stages} of one of the computing units.
In \Cref{fig:artistic_ico}, \circnum{6} annotates one of the top-level interconnect buses with individual pipeline stages visible.

\section{Conclusion \& Outlook}

In this work, we present a framework to embed images as top-metal ASIC art into existing \gls{gdsii} layouts and to render them artistically with ultra-high fidelity.
{\artistic} cannot only be used to generate outreach material for \gls{asic} projects to gain attention, share results, and secure funding, but also to build the foundation for scientific discussions concerning chip layouts.
With {\artistic} available open source, we hope to see many high-resolution \gls{asic} renders and posters emerging.

\printbibliography

\end{document}